# The Next Generation of Crystal Detectors

Ren-Yuan Zhu

California Institute of Technology, Pasadena, CA 91125, USA

March 29, 2013

**Introduction**

Crystal detectors have been used widely in high energy and nuclear physics experiments, medical instruments and homeland security applications. Novel crystal detectors are continuously being discovered and developed in academia and in industry. In high energy and nuclear physics experiments, total absorption electromagnetic calorimeters (ECAL) made of inorganic crystals are known for their superb energy resolution and detection efficiency for photon and electron measurements [1]. A crystal ECAL is thus the choice for those experiments where precision measurements of photons and electrons are crucial for their physics missions. Examples are the Crystal Ball NaI(Tl) ECAL, the L3 BGO ECAL and the BaBar CsI(Tl) ECAL in lepton colliders, the kTeV CsI ECAL and the CMS PWO ECAL in hadron colliders and the Fermi CsI(Tl) ECAL in space. For future HEP experiments at the energy and intensity frontiers, however, the crystal detectors used in the above mentioned ECALs are either not bright and fast enough, or not radiation hard enough. Crystal detectors have also been proposed to build a Homogeneous Hadron Calorimeter (HHCAL) to achieve unprecedented jet mass resolution by duel readout of both Cherenkov and scintillation light [2], where development of cost-effective crystal detectors is a crucial issue because of the huge crystal volume required [3]. This contribution discusses several R&D directions for the next generation of crystal detectors for future HEP experiments.

**Crystal detectors with excellent radiation hardness**:

With $10^{35}$ cm$^{-2}$s$^{-1}$ luminosity and 3,000 fb$^{-1}$ integrated luminosity the HL-LHC presents the most severe radiation environment, where up to 10,000 rad/h and $10^{15}$ hadrons/cm$^2$ are expected by the CMS forward calorimeter located at $\eta = 3$. Superb radiation hardness is required for any detector to be used at the HL-LHC. Fast timing is also required to reduce the pile-up effect. With high density (7.1 g/cm$^3$), fast decay time (40 ns) and superb radiation hardness against gamma-rays and charged hadrons cerium doped lutetium oxyorthosilicate (Lu$_2$SiO$_5$:Ce, LSO) [4] and lutetium yttrium oxyorthosilicate (Lu2$_{(1-x)}$Y$_{2x}$SiO$_5$:Ce, LYSO) [5] crystals are good candidate for future HEP experiments at both the energy and intensity frontiers. An LSO/LYSO crystal ECAL is base-lined for the Mu2e experiment [6] at the intensity frontier. An LSO/LYSO crystal based sampling ECAL is also proposed for the CMS forward calorimeter upgrade at the HL-LHC [7]. Because of its excellent radiation hardness and short light path an LSO/LYSO crystal based sampling ECAL is expected to provide a stable calorimeter for the CMS experiment at the HL-LHC.

R&D issues for crystal detectors with excellent radiation hardness include the following aspects.

- While the gamma-rays and protons induced radiation damage in crystal detectors has been investigated, only limited data are available for hadron induced damage in various crystal scintillators. A R&D program to investigate radiation damage effect induced by gamma-rays, charged hadrons as well as neutrons in various crystal detectors is needed to address this issue.

- LSO/LYSO crystals are relative expansive because of the high cost of raw material lutetium oxide and high melting point. Other fast scintillators (See Table 1 below) as well as fast scintillating glass and ceramics may also be considered. Development of fast and radiation hard crystal, glass and ceramics detectors is of general interest.

- For an LSO/LYSO based Shashlik ECAL radiation hard photo-detectors as well as wavelength sifters need also be developed.

**Crystal detectors with fast response time**:

Future HEP experiments at the energy and intensity frontiers require very fast crystals to cope with the unprecedented event rate [8].

Table 1 Basic Properties of Fast Crystal Scintillators

| | LSO/LYSO | GSO | YSO | CsI | $BaF_2$ | $CeF_3$ | $CeBr_3$ [1] | $LaCl_3$ | $LaBr_3$ | Plastic scintillator (BC 404) [2] |
|---|---|---|---|---|---|---|---|---|---|---|
| Density (g/cm$^3$) | 7.40 | 6.71 | 4.44 | 4.51 | 4.89 | 6.16 | 5.23 | 3.86 | 5.29 | 1.03 |
| Melting point (°C) | 2050 | 1950 | 1980 | 621 | 1280 | 1460 | 722 | 858 | 783 | 70[#] |
| Radiation Length (cm) | 1.14 | 1.38 | 3.11 | 1.86 | 2.03 | 1.70 | 1.96 | 2.81 | 1.88 | 42.54 |
| Molière Radius (cm) | 2.07 | 2.23 | 2.93 | 3.57 | 3.10 | 2.41 | 2.97 | 3.71 | 2.85 | 9.59 |
| Interaction Length (cm) | 20.9 | 22.2 | 27.9 | 39.3 | 30.7 | 23.2 | 31.5 | 37.6 | 30.4 | 78.8 |
| Z value | 64.8 | 57.9 | 33.3 | 54.0 | 51.6 | 50.8 | 45.6 | 47.3 | 45.6 | - |
| dE/dX (MeV/cm) | 9.55 | 8.88 | 6.56 | 5.56 | 6.52 | 8.42 | 6.65 | 5.27 | 6.90 | 2.02 |
| Emission Peak[a] (nm) | 420 | 430 | 420 | 420 / 310 | 300 / 220 | 340 / 300 | 371 | 335 | 356 | 408 |
| Refractive Index[b] | 1.82 | 1.85 | 1.80 | 1.95 | 1.50 | 1.62 | 1.9 | 1.9 | 1.9 | 1.58 |
| Relative Light Yield[a,c] | 100 | 45 | 76 | 4.2 / 1.3 | 42 / 4.8 | 8.6 | 141 | 15 / 49 | 153 | 35 |
| Decay Time[a] (ns) | 40 | 73 | 60 | 30 / 6 | 650 / 0.9 | 30 | 17 | 570 / 24 | 20 | 1.8 |
| d(LY)/dT [a,d] (%/°C) | -0.2 | -0.4 | -0.1 | -1.4 | -1.9 / 0.1 | ~0 | -0.1 | 0.1 | 0.2 | ~0 |

a. Top line: slow component, bottom line: fast component.
b. At the wavelength of the emission maximum.
c. Relative light yield normalized to the light yield of LSO
d. At room temperature (20°C)
#. Softening point

1. W. Drozdowski et al. *IEEE TRANS. NUCL. SCI*, VOL.55, NO.3 (2008) 1391-1396
   Chenliang Li et al, *Solid State Commun*, Volume 144, Issues 5–6 (2007),220–224
   http://scintillator.lbl.gov/
2. http://www.detectors.saint-gobain.com/Plastic-Scintillator.aspx
   http://pdg.lbl.gov/2008/AtomicNuclearProperties/HTML_PAGES/216.html

Table 1 above summarizes basic properties of fast crystal scintillators, including oxides (LSO/LYSO), halides ($BaF_2$, CsI and $CeF_3$) and the recently discovered bright and fast halides ($CeBr_3$, $LaCl_3$ and $LaBr_3$) as well as a typical plastic scintillator. All fast halides discovered recently are highly hygroscopic so that their application needs extra engineering work. Among the non-hygroscopic crystal scintillators listed in Table 1, $BaF_2$ crystals with a sub-nanosecond fast decay component produce the highest scintillation photon yield in the $1^{st}$ nano-second, just less than highly hygroscopic $CeBr_3$ and $LaBr_3$. This fast component provides a solid foundation for an extreme fast ECAL made by $BaF_2$ crystals. In addition to the good energy resolution and fast response time the photon direction measurements are also important for neutral pion identification through its two photon decays. Most crystal ECAL built so far has no longitudinal segmentation because of the technical difficulty of imbedding readout device in a total absorption calorimeter. One exception is the GLAST ECAL which is a CsI(Tl) strips based hodoscope [9]. With compact readout devices developed in the last decade this issue is less relevant now. A well designed segmented ECAL with fast crystal detectors may provide excellent energy resolution, fast timing and fine photon angular resolution, so would serve well for the future HEP experiments in general and those at the intensity frontier in particular.

R&D issues for fast crystal detectors include the following aspects.

- The slow scintillation component in $BaF_2$ needs to be suppressed by selective doping and/or selective readout by using solar blind photo-detector. The radiation harness of $BaF_2$ crystals needs to be further improved.

- Development of novel fast crystal detectors with sub-nanosecond decay time, e.g. YAP:Yb [10] etc., is of general interest.

- For a segmented crystal calorimeter with photon pointing resolution R&D on detector design and compact readout devices are required.

**Crystal detectors for excellent jet mass reconstruction**

The next generation of collider detectors will emphasize precision for all sub-detector systems. One of the benchmarks is to distinguish W and Z bosons in their hadronic decay mode. Excellent jet mass resolution is required for future lepton colliders at the energy frontier. Following successful experiences with crystal ECAL, a novel HHCAL detector concept was proposed to achieve unprecedented jet mass resolution by duel readout of both Cherenkov and scintillation light. The HHCAL detector concept includes both electromagnetic and hadronic parts, with separate readout for the Cherenkov and scintillation light and using their correlation to obtain superior hadronic energy resolution. It is a total absorption device, so its energy resolution is not limited by sampling fluctuations. It also has no structural boundary between the ECAL and HCAL, so it does not suffer from the effect of dead material in the middle of hadronic showers. With the dual-readout approach large fluctuations in the determination of the electromagnetic fraction of a hadronic shower can be reduced. The missing nuclear binding energy can then be corrected shower by shower, resulting in a good energy resolution for hadronic jets. Such a calorimeter would thus provide excellent energy resolution for photons, electrons and hadronic jets.

R&D issues for crystal detectors for the HHCAL detector concept include the following aspects:

- Development of cost-effective crystal detectors is crucial because of the unprecedented volume (70 to 100 m$^3$) foreseen for an HHCAL. The material of choice must be dense, cost-effective, UV transparent (for effective collection of the Cherenkov light) and allow for a clear discrimination between the Cherenkov and scintillation light. The preferred scintillation light thus is at a longer wavelength, and not necessarily bright or fast.

Table 2. Basic properties of candidate crystal detectors for HHCAL

| Parameters | $Bi_4Ge_3O_{12}$ (BGO) | $PbWO_4$ (PWO) | $PbF_2$ | PbClF | $Bi_4Si_3O_{12}$ (BSO) |
|---|---|---|---|---|---|
| ρ (g/cm$^3$) | 7.13 | 8.29 | 7.77 | 7.11 | 6.8 |
| $\lambda_I$ (cm) | 22.8 | 20.7 | 21.0 | 24.3 | 23.1 |
| n @ $\lambda_{max}$ | 2.15 | 2.20 | 1.82 | 2.15 | 2.06 |
| $\tau_{decay}$ (ns) | 300 | 30/10 | ? | 30 | 100 |
| $\lambda_{max}$ (nm) | 480 | 425/420 | ? | 420 | 470 |
| Cut-off λ (nm) | 310 | 350 | 250 | 280 | 300 |
| Light Output (%) | 100 | 1.4/0.37 | ? | 17 | 20 |
| Melting point (°C) | 1050 | 1123 | 842 | 608 | 1030 |
| Raw Material Cost (%) | 100 | 49 | 29 | 29 | 47 |

Table 2 above compares basic properties of candidate cost-effective crystal detectors for the HHCAL detector concept, such as PWO, scintillating $PbF_2$ [3], PbFCl [11] and BSO. Scintillating glasses and ceramics are also attractive for this detector concept.

- Development of compact photo-detectors for scintillation and Cherenkov light and design study for segmented crystal calorimeter with compact readout.

**Summary**

Crystal detectors have been used widely for decades in high energy and nuclear physics experiments, medical instruments and homeland security applications. Future HEP experiments require bright and fast crystal detectors with excellent radiation hardness. Cost-effectiveness is also a crucial issue for crystal detectors to be used in a large volume. Crystal detectors are also an area of fast development. New materials are continuously being found each year. To face these challenges a thorough R&D program is required to investigate and develop crystal detectors for future HEP experiments in all frontiers.